\documentclass[12pt,leqno]{article}
\usepackage{amsmath,amsthm}

\def\init{\setcounter{equation}{0}}
\newtheorem{theorem}{Theorem}[section]

\newcommand{\R}{{\bf R}}

\newtheorem{pro}{Proposition}[section]

\newcommand{\e}{{\varepsilon}}

\title{Inverse hyperbolic problems and optical black holes.
\author{G.Eskin, \ \ \  Department of Mathematics, UCLA,\\ Los Angeles,
CA 90095-1555, USA. \ E-mail: eskin@math.ucla.edu}
}

\begin{document}

\maketitle
\flushright{To the  memory of Leonid Romanovich Volevich
\ 
\\
\
\\
\ }

\begin{abstract}
In this paper we give a more geometrical formulation of the main theorem in [E1] on
the inverse problem 
for the second order hyperbolic equation of general form with coefficients independent
of the time variable.  We apply this theorem to the inverse problem for the equation of
the propagation of light in a moving medium (the Gordon equation).  Then we study the existence of 
black and white holes for the general hyperbolic and for the Gordon equation and we discuss 
the impact of this phenomenon on the inverse problems. 
\end{abstract}

\section{The refined formulation of the main theorem.}
\label{section 1}
\init
 
Let $\Omega$  be a smooth bounded domain in $\R^n$.  Consider a second order hyperbolic equation
in a cylinder $\Omega\times(-\infty,+\infty)$:
\begin{equation}                                               \label{eq:1.1}
\sum_{j,k=0}^n\frac{1}{\sqrt{(-1)^ng(x)}}\frac{\partial}{\partial x_j}\left(\sqrt{(-1)^ng(x)}g^{jk}(x)
\frac{\partial u(x_0,x)}{\partial x_k}\right)=0,
\end{equation}
where
$x=(x_1,...,x_n)\in \Omega,\ x_0\in \R$ is the time variable,  $g(x)=(\det[g^{jk}(x)]_{j,k=0}^n)^{-1}$
We assume that $g^{jk}(x)=g^{kj}(x)$ 
are real-valued smooth function in $C^\infty(\overline{\Omega})$ independent of $x_0$.   

The hyperbolicity of (\ref{eq:1.1})  means that the quadratic form $\sum_{j,k=0}^ng^{jk}(x)\xi_j\xi_k$
has one positive and $n$  negative eigenvalues,  $\forall x\in \overline{\Omega}$.  We assume in
addition that
\begin{equation}                                               \label{eq:1.2}
g^{00}(x)>0,\ \ x\in \overline{\Omega},
\end{equation}
i.e. $(1,0,...,0)$ is not a characteristic direction,
and that 
\begin{equation}                                              \label{eq:1.3}
\sum_{j,k=1}^ng^{jk}(x)\xi_j\xi_k<0\ \ \ \ \mbox{for\ \ }\forall (\xi_1,...,\xi_n) \neq
(0,...,0),\ \forall x\in \overline{\Omega},
\end{equation}
i.e.  the quadratic form  (\ref{eq:1.3})   is negative definite.
Note that (\ref{eq:1.3}) equivalent to the condition that
\begin{equation}                                               \label{eq:1.4}
\mbox{any direction \ \ } (0,\xi)\ \ 
\mbox{is not characteristic for\ \ } \forall x\in\overline{\Omega}.
\end{equation}
We shall  give later another equivalent characterization of the condition (\ref{eq:1.3}).

We shall study the initial-boundary value problem for the equation (\ref{eq:1.1}) in 
$\Omega\times\R$
\begin{equation}                                           \label{eq:1.5}    
u(x_0,x)=0\ \ \ \ \mbox{for}\ \ x_0<<0,\ \ x\in \Omega,
\end{equation}
\begin{equation}                                          \label{eq:1.6}
u|_{\partial\Omega\times\R}=f,
\end{equation}
where
$f(x_0,x')$  has a compact support in $\partial\Omega\times\R$.

Consider the Dirichlet-to-Neumann (DN) operator
\begin{equation}                                            \label{eq:1.7}
\Lambda f=\sum_{j,k=0}^ng^{jk}(x)\frac{\partial u}{\partial x_j}\nu_k(x)
\left.|\sum_{p,r=1}^ng^{jk}(x)\nu_p\nu_r|^{-\frac{1}{2}}\right|_{\partial\Omega\times\R},
\end{equation}
where
$\nu_0=0,(\nu_1,...,\nu_n)$  is the unit outward normal to $\partial\Omega\subset \R^n$  
with respect to the Euclidean metric,  $u(x_0,x)$  is the solution of 
(\ref{eq:1.1}), (\ref{eq:1.5}),  (\ref{eq:1.6}).

Let $\Gamma_0\subset\partial\Omega$  be an open subset of $\partial\Omega$.
We shall say that the DN operator $\Lambda$  is given on $\Gamma_0\times(0,T_0)$  if
we know $\Lambda f|_{\Gamma_0\times(0,T_0)}$  for any $f$  with the support in
$\overline{\Gamma}_0\times[0,T_0]$.

Let $X$  be a closed compact subset of $\partial\Omega\times(-\infty,+\infty)$.
For each distribution $f$ on $\partial\Omega\times(-\infty,+\infty),\ \ \mbox{supp\ }f\subset X$,  
denote by $D_+(f)$ the support in $\overline{\Omega}\times(-\infty,+\infty)$  of the solution
of the initial-boundary value prolem for (\ref{eq:1.1})  with $u=0$  for $x_0<<0,
\ x\in \Omega,\ u|_{\partial\Omega\times(-\infty,+\infty)}=f$.

We define the forward domain of influence of $X$  as the closure of the set 
$\cup_{\mbox{supp\ }f\subset X} D_+(f)$,  
where the union is taken over all $f$  on $\partial\Omega\times(-\infty,+\infty)$  with supports
in $X$.

We shall give a geometrical description of the set $D_+(X)$.  Let $[g_{jk}(x)]_{j,k=0}^n=
([g^{jk}]_{j,k=0}^n)^{-1}$,   i.e.  $[g_{jk}(x)]_{j,k=0}^n$  is the pseudo-Riemannian metric tensor.
We shall say that 
$\gamma(s)=(x_0(s),x_1(s),...,x_n(s)),\ s\in [0,s_0]$  is a forward
time-like path (ray)  if $\gamma(s)$  is continuous and piece-wise smooth and each smooth segment
$\gamma^{(i)}(s)$  of $\gamma(s)$  satisfies
\begin{equation}                                            \label{eq:1.8}
\sum_{j,k=0}^ng_{jk}(x(s))\frac{dx_j}{ds}\frac{dx_k}{ds}>0,\ \ s_i\leq s\leq s_{i+1},
\end{equation}
$\frac{dx_0}{ds}>0$.
Then $D_+(X)$  is the closure in $\overline{\Omega}\times(-\infty,+\infty)$
of the union of all forward time-like paths (rays)  starting on $\overline{X}$.  
Analogously $D_-(X)$  is called a domain of dependence of $X$  if $D_-(X)$  is the union of
all backward time-like rays starting at $\overline{X}$.
We 
say that piecewise smooth $\gamma(s),\ s\geq 0$,  is a backward time-like ray if (\ref{eq:1.8})
holds for each smooth segment,  and $\frac{dx_0}{ds}<0$  on $\gamma^{(i)}(s),\ \forall s \in [s_i,s_{i+1}]$.
Note that the vertical ray $x_0=s,\ x=x^{(0)},\ s\geq 0$,  is time-like iff $g_{00}(x^{(0)})>0.$

\begin{pro}                                           \label{pro:1.1}
Condition (\ref{eq:1.3})  holds iff $g_{00}(x)>0$,  i.e.  the ray $x_0=s, \ x\in \overline{\Omega},\ 
s\geq 0$,  is time-like.
\end{pro}
{\bf Proof:}  We have 
$$
g_{00}(x)=\frac{\det[g^{jk}]_{j,k=1}^n}{g^{-1}(x)}.
$$
Note that 
$$
\sum_{j,k=0}^ng^{jk}(x)\xi_j\xi_k=\left(\sqrt{g^{00}(x)}\xi_0
+\sum_{j=1}^n \frac{g^{0j}(x)}{\sqrt{g_{00}(x)}}\xi_j\right)^2
$$
$$
-\left(\sum_{j=1}^n\frac{g^{0j}(x)}{\sqrt{g_{00}(x)}}\xi_j\right)^2+\sum_{j,k=1}^ng^{jk}(x)\xi_j\xi_k.
$$

Since the quadratic form $\sum_{j,k=0}^ng^{jk}\xi_j\xi_k$  has the signature 
$(+1,-1,...,-1)$   the quadratic form 
$$
-\left(\sum_{j=1}^n\frac{g^{jk}(x)}{\sqrt{g_{00}(x)}}\xi_j\right)^2
+\sum_{j,k=1}^n g^{jk}(x)\xi_j\xi_k
$$
is negative definite.  Therefore 
$\sum_{j,k=1}^n g^{jk}(x)\xi_j\xi_k$
either has only negative eigenvalues,  or it has one zero eigenvalue,  or it has 
one positive eigenvalue.  
In the first case $\ \mbox{sgn}(\det[g^{jk}]_{j,k=1}^n)=(-1)^n,$ in the second case
$\det[g^{jk}(x)]_{j,k=1}^n=0$,
and in the third case  $\ \mbox{sgn}(\det[g^{jk}]_{j,k=1}^n)=(-1)^{n-1}$.
Since 
$\ \mbox{sgn}\ g(x)=(-1)^{n}$
we get that $g_{00}(x)>0$ iff condition (\ref{eq:1.3}) holds.
\qed

Let $T_+$  be the smallest number  such that 
$D_+(\overline{\Gamma}_0\times\{x_0=0\})\supset
\overline{\Omega}\times \{x_0=T_+\}$.
Analoguously let $T_-$  be the smallest number such that 
$D_-(\overline{\Gamma}_0\times\{x_0=T_-\}) 
\supset\overline{\Omega}\times\{x_0=0\}.$

We shall require that
\begin{equation}                                            \label{eq:1.9}
T_0>T_++T_-.
\end{equation}
Consider the following change of variables in $\overline{\Omega}\times(-\infty,+\infty)$:
\begin{eqnarray}                                       \label{eq:1.10}
y_0=x_0+a(x),
\\
\nonumber
y=\varphi(x),
\end{eqnarray}
where $y=\varphi(x)$  is a diffeomorphism of $\overline{\Omega}$ onto
$\overline{\tilde{\Omega}},\ a(x)\in C^\infty(\overline{\Omega})$,
\begin{eqnarray}                                       \label{eq:1.11}
\varphi(x)=I\ \ \mbox{on}\ \ \overline{\Gamma}_0\subset\partial\Omega,
\\
\nonumber
a(x)=0\ \ \mbox{on}\ \ \overline{\Gamma}_0.
\end{eqnarray}
Note that if 
$\tilde{u}(y_0,y)=u(x_0,x)$,  where $(y_0,y)$  and $(x_0,x)$ are related
by (\ref{eq:1.10}),  (\ref{eq:1.11}),
then $\tilde{u}(y_0,y)$  satisfies an equation of the form
\begin{equation}                                               \label{eq:1.12}
\sum_{j,k=0}^n\frac{1}{\sqrt{(-1)^n\tilde{g}(y)}}
\frac{\partial}{\partial y_j}\left(\sqrt{(-1)^n\tilde{g}(y)}\tilde{g}^{jk}(y)
\frac{\partial \tilde{u}(y_0,y)}{\partial y_k}\right)=0
\end{equation}
in $\tilde{\Omega}\times(-\infty,+\infty)$  with initial and boundary conditions
\begin{equation}                                              \label{eq:1.13}
\tilde{u}(y_0,y)=0\ \ \mbox{for}\ \ y_0<<0,\ \ y\in\tilde{\Omega},
\end{equation}
\begin{equation}                                              \label{eq:1.14}
\tilde{u}|_{\partial\tilde{\Omega}\times\R}=f(y_0,y'),
\end{equation}
where  $f(x_0,x')=f(y_0,y')$  on $\Gamma_0\times(-\infty,+\infty)$.

Let $\tilde{\Lambda}$  be the DN operator on $\Gamma_0\times(-\infty,+\infty)$
corresponding to (\ref{eq:1.12}),  (\ref{eq:1.13}),  (\ref{eq:1.14}).

The following theorem holds (c.f. [E1], Theorem 2.3)

\begin{theorem}                                     \label{theo:1.1}
Let  (\ref{eq:1.1}),  (\ref{eq:1.5}),  (\ref{eq:1.6})  and
(\ref{eq:1.12}),  (\ref{eq:1.13}),  (\ref{eq:1.14})  be two hyperbolic initial-boundary value
problems  in $\Omega\times\R$ and $\tilde{\Omega}\times\R$,  respectively,
and $\partial\Omega\cap\partial\tilde{\Omega}\supset\Gamma_0$.
Suppose conditions (\ref{eq:1.2}) and (\ref{eq:1.3}) are satisfied for (\ref{eq:1.1}) and 
(\ref{eq:1.12}).  Suppose $\mbox{supp\ }f\subset\overline{\Gamma}_0\times[0,T_0]$ in
(\ref{eq:1.6}) and (\ref{eq:1.14}).  If the DN operators $\Lambda$ and $\tilde{\Lambda}$
are equal on $\Gamma_0\times(0,T_0)$  and if $T_0>T_++T_-$ (c.f. (\ref{eq:1.9}))  then there
exists a change of variables (\ref{eq:1.10}),  (\ref{eq:1.11})
such that
\begin{equation}                                \label{eq:1.15}
[\tilde{g}^{jk}(y)]_{j,k=0}^n=J(x)[g^{jk}(x)]_{j,k=0}^nJ^T(x),
\end{equation}
where $y_0=x_0+a(x),\ y=\varphi(x),$
\begin{equation}                                 \label{eq:1.16}
J(x)=\left[ \begin{array}{ll}  1  &   a_x(x)\\
a_x^T(x)   & \frac{\mathcal{D}\varphi(x)}{\mathcal{D}x}
\end{array}
\right]
\end{equation}
is the Jacobi matrix of (\ref{eq:1.10}).
\end{theorem}

Note that 
condition (\ref{eq:1.9}) is required for (\ref{eq:1.1}) only.
The condition
(\ref{eq:1.15}) is equivalent to 
$$
[\tilde{g}^{jk}(y)]^{-1}=(J^T(x))^{-1}[g^{jk}(x)]^{-1}J^{-1}(x),
$$
or
\begin{equation}                               \label{eq:1.17}
[g_{jk}(x)]_{j,k=0}^n=J^T(x)[\tilde{g}_{jk}(y)]_{j,k=0}^nJ(x).
\end{equation}
The equality (\ref{eq:1.17}) can be rewritten as the equality of differential forms
\begin{equation}                                \label{eq:1.18}
\sum_{j,k=0}^ng_{jk}(x)dx_jdx_k=\sum_{j,k=1}^n\tilde{g}_{jk}(y)dy_jdy_k.
\end{equation}
The proof of Theorem \ref{theo:1.1} is the same as the proof of Theorem 2.3 in [E1] (see also [E2]).

\section{The inverse problem for the Gordon equation.}
\label{section 2}
\init
 
Consider the equation of the propagation of light 
in a moving medium.  Let $w(x)=(w_1(x),w_2(x),w_3(x))$   be the velocity of
flow and let $(v^{(0)},v^{(1)},v^{(2)},v^{(3)})$  be the corresponding four-velocity 
vector of flow:
$$
v^{(0)}=
\left(1-\frac{|w|^2}{c^2}\right)^{-\frac{1}{2}},\ \ \
v^{(j)}=
\left(1-\frac{|w|^2}{c^2}\right)^{-\frac{1}{2}}\frac{w_j(x)}{c},\ \ 1\leq j\leq 3,
$$
where $c$ is the speed of light in the vacuum (c.f.  [G], [NVV], [LP]).  The equation
for the propagation of light was found in 1923 by Gordon (c.f. [G]) and it has the form
(\ref{eq:1.1})  when $g^{jk}(x),\ 0\leq j,k\leq n,\ n=3$,
are
\begin{equation}                                      \label{eq:2.1}
g^{jk}(x)=\eta^{jk}+(n^2(x)-1)v^j(x)v^k(x),
\end{equation}
$[\eta_{jk}]=[\eta^{jk}]^{-1}$  is the Lorentz metric tensor:
$\eta^{jk}=0$ when $j\neq k,\ \eta^{00}=1,\ \eta^{jj}=-1,\ 1\leq j\leq 3,\ 
n(x)=\sqrt{\e(x)\mu(x)}$  is the refraction index.
Note that $[g_{jk}(x)]=[g^{jk}]^{-1}$  has the form (c.f. [LP]):
\begin{equation}                                      \label{eq:2.2}
g_{jk}(x)=\eta_{jk}+(n^{-2}(x)-1)v_jv_k,\ \ 0\leq j,k\leq n,
\end{equation}
where $v_0=v^0,\ v_j=-v^j,\ 1\leq j\leq n$.

The metric (\ref{eq:2.2}) is called the Gordon metric
and the corresponding equation of the form (\ref{eq:1.1}) is called 
the Gordon equation.
We shall consider the Gordon equation in $\Omega\times(-\infty,+\infty)$,  where 
$\Omega$  has the form  
$\Omega=\Omega_0\setminus\cup_{j=1}^m\overline{\Omega}_j,\ \Omega_0$  is 
diffeomorphic to a ball,\ $\Omega_j$  are smooth domains,
$\overline{\Omega}_j\cap\overline{\Omega}_k=\emptyset$  when 
$j\neq k$  and $\cup_{j=1}^m\overline{\Omega}_j\subset\Omega$.
Domains $\Omega_j$ are called obstacles.  

We shall
consider the following initial-bundary value problem for the Gordon equation:
\begin{equation}                                       \label{eq:2.3}
u(x_0,x)=0\ \ \ \mbox{for}\ \ x_0<<0,\ x\in\Omega,
\end{equation}
\begin{equation}                                       \label{eq:2.4}
u(x_0,x)|_{\partial\Omega_j\times\R}=0,\ \ 1\leq j\leq\ m,
\end{equation}
\begin{equation}                                       \label{eq:2.5}
u(x_0,x)|_{\partial\Omega_0\times\R}=f(x_0,x),
\end{equation}
i.e.
$\Gamma_0=\partial\Omega_0$  where $\Gamma_0$  is the same as in Theorem \ref{theo:1.1}.
One can consider also the case when on some or all obstacles $\Omega_j$  the zero
Dirichlet boundary condition is replaced by the zero Neumann boundary condition.

Note that the condition (\ref{eq:1.2}) always holds for the Gordon equation
\begin{equation}                                        \label{eq:2.6}
g^{00}=1+(n^2-1)(v^0)^2>0.
\end{equation}
The condition (\ref{eq:1.3}) holds iff
\begin{equation}                                        \label{eq:2.7}
|w(x)|^2<\frac{c^2}{n^2(x)}.
\end{equation}
To prove (\ref{eq:2.7})
note that the principal symbol of the Gordon equation is
$$
\xi_0^2-|\xi|^2+(n^2-1)\left(\sum_{j=0}^nv^j\xi_j\right)^2.
$$
Therefore the quadratic form
\begin{equation}                                   \label{eq:2.8}
-|\xi|^2+(n^2-1)\left(\sum_{j=1}^nv^j\xi_j\right)^2
\end{equation}
is negative definite iff $(n^2-1)|v|^2<1$.
Since $v^0=\left(1-\frac{|w|^2}{c^2}\right)^{-\frac{1}{2}},\ v^j=
\left(1-\frac{|w|^2}{c^2}\right)^{-\frac{1}{2}}\frac{w_j}{c},\ 1\leq j\leq n$, we get 
(\ref{eq:2.7}).

We shall study the inverse problem for the Gordon equation.  The case of slowly moving 
medium (c.f. [LP]) was considered in [E1].  The Dirichlet-to-Neumann operator corresponding
to the Gordon equation has the form (c.f. (\ref{eq:1.7})):
%\begin{equation}                                             \label{eq:2.9}
$$
\Lambda f=\sum_{j,k=0}^n[\eta^{jk}+(n^2(x)-1)v^jv^k]\frac{\partial u}{\partial x_j}\nu_k
\left.|-1+(n^2-1)(\sum_{r=1}^nv^r\nu_r)^2|^{-\frac{1}{2}}
\right|_{\partial\Omega_0\times\R},
$$
%\end{equation}
where
$\nu(x)=(\nu_1,...,\nu_n)$  is the unit outward normal vector to $\partial\Omega_0,\ \nu_0=0,\ 
\left.u\right|_{\partial\Omega_0\times\R}=f$.

We shall impose some restriction on the flow $w(x)=(w_1(x),...,w_n(x))$.

Let $x=x(s),\ s\in[0,s_0]$  be a trajectory of the flow,  i.e.  $\frac{dx}{ds}=
w(x(s)),\ w(x(s))\neq 0,\ s\in [0,s_0]$.  Consider the trajectories that either start and end on
$\partial\Omega_0$  or are closed curves in $\Omega$.  Our assumption is that 
\begin{eqnarray}                                   \label{eq:2.9}
\mbox{The union of all such trajectories}
\ \
%\nonumber
\mbox{is a dense set in \ }\ \Omega.
\end{eqnarray}

\begin{theorem}                                   \label{theo:2.1}
Let $[g_{jk}(x)]_{j,k=0}^n$ and $[\tilde{g}_{jk}(y)]_{j,k=0}^n$
be two Gordon metrics in domains $\Omega$  and $\tilde{\Omega}$,  respectively.
Consider two initial-boundary value problems
(\ref{eq:1.1}),  (\ref{eq:1.5}),  (\ref{eq:1.6})  and
(\ref{eq:1.12}),  (\ref{eq:1.13}),  (\ref{eq:1.14}) corresponding to the metrics 
$[g_{jk}(x)]_{j,k=0}^n$ and $[\tilde{g}_{jk}(y)]_{j,k=0}^n$,
respectively.
Assume that that the condition (\ref{eq:2.7}) holds for both metrics and assume that
the refraction coefficients $n^2(x)$  and $\tilde{n}^2(y)$ are constant.  Assume also that
the flow $w(x)$ satisfies the condition (\ref{eq:2.9}).
If the DN operators $\Lambda$ and $\tilde{\Lambda}$ corresponding to 
$[g_{jk}(x)]_{j,k=0}^n$ and $[\tilde{g}_{jk}(y)]_{j,k=0}^n$
are equal on $\partial\Omega_0\times (0,T_0)$,  where $T_0$  satisfies (\ref{eq:1.9})
for equation (\ref{eq:1.1}),  then we have
$$
n^2=\tilde{n}^2,\ \ \Omega=\tilde{\Omega},
$$
and the flows $w(x)$  and $\tilde{w}(x)$ are equal.
\end{theorem}

{\bf Proof:}
Analogously to the Remark 2.2 in [E1]  we can find the symbol of the DN operator
in the "elliptic" region (c.f. [E1])  and retrieve  $\left.n^2(x)\right|_{\partial\Omega_0}$.
Since $\Lambda=\tilde{\Lambda}$ on  $\partial\Omega_0\times(0,T_0)$  we  get that 
$\left.n^2(x)\right|_{\partial\Omega_0}=\left.\tilde{n}^2(x)\right|_{\partial\Omega_0}$.
Therefore $n^2=\tilde{n}^2$  in $\overline{\Omega}$  since we assume that 
$n^2$ and $\tilde{n}^2$ do not depend on $x$.  Applying  Theorem \ref{theo:1.1}
to the case  of the Gordon equation we get that there exists a change of variables (\ref{eq:1.10})
such that $a(x)=0$  on $\partial\Omega_0,\ \varphi(x)=I$  on $\partial\Omega_0$  and
(\ref{eq:1.15}),  (\ref{eq:1.18}) hold where
\begin{eqnarray}                                            \label{eq:2.10}
g^{jk}(x)=\eta^{jk}+(n^2-1)v^j(x)v^k(x),\ \ 0\leq j,k\leq n,
\\
\nonumber
\tilde{g}^{jk}(y)=\eta^{jk}+(n^2-1)\tilde{v}^j(y)\tilde{v}^k(y),\ \ 0\leq j,k\leq n.
\end{eqnarray}
We have from (\ref{eq:1.10})
\begin{eqnarray}                                            \label{eq:2.11}
dy_0=dx_0+\sum_{k=1}^n a_{x_k}(x)dx_k,
\\
\nonumber
dy_j=\sum_{k=1}^n\varphi_{jx_k}(x)dx_k,\ \ 1\leq k\leq n.
\end{eqnarray}
Substituting (\ref{eq:2.11}) into (\ref{eq:1.18}) and taking into account that 
$dx_0,dx_1,...,dx_n$  are arbitrary we get
$$
g_{00}(x)=\tilde{g}_{00}(y).
$$
Therefore (c.f. (\ref{eq:2.2}))
\begin{equation}                                      \label{eq:2.12}
1+(n^{-2}-1)v_0^2(x)=1+(\tilde{n}^{-2}-1)\tilde{v}_0^2(y).
\end{equation}
Note that (\ref{eq:1.15}) is equivalent to
\begin{equation}                                      \label{eq:2.13}
\tilde{g}^{jk}(y)=\sum_{p,r=0}^ng^{pr}(x)\varphi_{jx_p}\varphi_{kx_r},
\end{equation}
where $\varphi_0=x_0+a(x)$.  In particular,
\begin{equation}                                         \label{eq:2.14}
\tilde{g}^{00}(y)=g^{00}(x)+2\sum_{p=1}^ng^{p0}(x)a_{x_p}(x)+
\sum_{p,r=1}^ng^{pr}(x)a_{x_p}a_{x_r}(x).
\end{equation}
In the case of Gordon metrics we have 
\begin{eqnarray}                                        \label{eq:2.15}
\
\\
\nonumber
1+(\tilde{n}^2-1)(\tilde{v}^0(y))^2
\ \ \ \ \ \ \ \ \ \ \ \ \ \ \ \ \ \ \ \ \ \ \ \ \ \ \ \ \ \ \ \ \ \ \ \ 
\ \ \ \ \ \ \  
\\
\nonumber
=1+(n^2-1)(v^0(x))^2
+2\sum_{p=1}^n(n^2-1)v^0(x)v^p(x)a_{x_p}(x)
\\
\nonumber
-
\sum_{p=1}^n(a_{x_{x_p}}(x))^2
+\sum_{p,r=1}^n(n^2-1)v^p(x)v^r(x)a_{x_p}(x)a_{x_r}(x).
\end{eqnarray}
Denote 
$$
|\nabla a(x)|^2=\sum_{p=1}^na_{x_p}^2(x),\ \ \
a_0(x)=\frac{1}{c}\sum_{p=1}^nw_p(x)a_{x_p}(x).
$$
Since $n=\tilde{n}$  and since (\ref{eq:2.12}) implies that $v^0=\tilde{v}^0$
we get
\begin{equation}                                \label{eq:2.16}
a_0^2+2a_0(x)-\frac{|\nabla a|^2}{(v^0)^2(n^2-1)}=0.
\end{equation}
It follows from (\ref{eq:2.16}) that either
\begin{equation}                               \label{eq:2.17}
a_0=-1-\sqrt{1+\frac{|\nabla a|^2}{(v^0)^2(n^2-1)}},
\end{equation}
or
\begin{equation}                               \label{eq:2.18}
a_0=-1+\sqrt{1+\frac{|\nabla a|^2}{(v^0)^2(n^2-1)}}.
\end{equation}
Let $x=x(s)$  be a trajectory of the flow,  i.e.
$$
\frac{dx_k}{ds}=w_k(x(s)),\ \ 1\leq k\leq n, \ \ 0\leq s\leq 1.
$$
Then 
$$
\frac{da(x(s))}{ds}=\sum_{k=1}^na_{x_k}(x(s))\frac{dx_k}{ds}=ca_0(x(s)),
$$
i.e. $ca_0(x(s))$  is the derivative of $a(x(s))$  along the trajectory of the flow.

Suppose $x=x(s),\ 0\leq s\leq 1,$  is a tragectory that starts and
ends on $\partial\Omega_0$.  Therefore $a(x(0))=a(x(1))=0$ since $a(x)=0$  
on $\partial\Omega_0$.  Then
$a_0(x(s))$ can not satisfy (\ref{eq:2.17})
since $\frac{d a(x(s))}{ds}=ca_0(x(s))<0$  on $[0,1]$.  Therefore 
$\frac{d a(x(s))}{ds}=ca_0(x(s))$ satisfies the equation (\ref{eq:2.18}),  i.e. 
$\frac{d a(x(s))}{ds}\geq 0$ on $[0,1]$.  Since $a(x(0))=a(x(1))=0$
we must have $a(x(s))= 0$  and $\nabla a(x(s))=0$  on $[0,1]$.
In the case when $x=x(s),\ 0\leq s\leq 1,$  is a closed trajectory,  i.e.  
$x(0)=x(1),\ a(x(s))$ can not again satisfy (\ref{eq:2.17})   since 
$\frac{da(x(s))}{ds}<0$  on $[0,1]$   and $a(x(s))$  satisfies (\ref{eq:2.18}) only
when $\nabla a(x(s))=0$  and $a(x(s))=\mbox{const}$  on $[0,1]$.

Since the condition (\ref{eq:2.9}) holds and since $a|_{\partial\Omega}=0$
we get that  $a=0$  in $\overline{\Omega}$.  In fact  if $a(x^{(0)})\neq 0$  then
$a(x^{(0)})\neq 0$  in some neighborhood  $U_0$  of $x^{(0)}$.   Since
$\nabla a(x)=0$  in $U_0$  we get that $a(x)=a(x^{(0)})$  in $U_0$,  i.e.
the set $a(x)=a(x^{(0)})$  is simultaneously open and close.
Since $a=0$  on $\partial\Omega_0$  we get that 
$a(x^{(0)})=0$,  and this is a contradiction.

Applying (\ref{eq:2.13}) for $k=0,\ 1\leq j\leq n,$  and for $1\leq j,k\leq n$
and taking into account that $a\equiv 0$  we get
\begin{equation}                                             \label{eq:2.19}
(\tilde{n}^2-1)\tilde{v}^0(y)\tilde{v}^j(y)=(n^2-1)\sum_{p=1}^nv^0(x)v^p(x)\varphi_{jx_p}(x),
\end{equation}
\begin{eqnarray}                                            \label{eq:2.20}
-\eta^{jk}+(\tilde{n}^2-1)\tilde{v}^j(y)\tilde{v}^k(y)
=
-\sum_{p=1}^n\varphi_{jx_p}\varphi_{kx_p}
\\
\nonumber
+
\sum_{p,r=1}^n(n^2-1)v^p(x)v^r(x)\varphi_{jx_p}\varphi_{kx_r},\ \ \ 1\leq j,k\leq n.
\end{eqnarray}
Using again that $\tilde{n}=n$   and $v^0=\tilde{v}^0$  we get from (\ref{eq:2.19})
\begin{equation}                                           \label{eq:2.21}
\tilde{v}^j(y)=\sum_{p=1}^nv^p(x)\varphi_{jx_p}(x),\ \ 1\leq j\leq n,\ \tilde{v}^0(y)=v^0(x).
\end{equation}
Substituting (\ref{eq:2.21}) into (\ref{eq:2.20})  we obtain
\begin{equation}                                             \label{eq:2.22}
\eta^{jk}=\sum_{p=1}^n\varphi_{jx_p}(x)\varphi_{kx_p}(x),\ \ 1\leq j,k\leq n.
\end{equation}
In particular,  we have
\begin{equation}                                        \label{eq:2.23}
\sum_{p=1}^n\varphi_{jx_p}^2(x)=1,\ \ \varphi_j|_{\partial\Omega_0}=x_j,\ \ 1\leq j\leq n.
\end{equation}
The Cauchy problem (\ref{eq:2.23})  has a unique solution in 
$\overline{\Omega}:\varphi_j=x_j,\ 1\leq j\leq n.$
Therefore $y=\varphi(x)=x$  in $\overline{\Omega}$.  This implies that $\tilde{\Omega}=\Omega$.  Also
$\varphi_j(x)=x_j$  implies (c.f. (\ref{eq:2.21})) that 
$\tilde{v}^j(x)=v^j(x),\ 1\leq j\leq n$.

Therefore $\tilde{w}_j(x)=w_j(x),\ 1\leq j\leq n$.
\qed

\section{Optical black holes.}
\label{section 3}
\init

In this section we explore the situation when the condition (\ref{eq:1.3}) is not
satisfied.

Denote by $S$  the surface in $\Omega$  given by the equation $\Delta(x)=0$  where
\begin{equation}                                \label{eq:3.1}
\Delta(x)=\det[g^{jk}(x)]_{j,k=1}^n.
\end{equation}
We assume that $S$  is a smooth and closed surface.
Denote by $\Omega_{ext}$ the exterior of $S$ and by $\Omega_{int}$ the interior 
of $S$.  We assume that $\Delta >0$  in $\Omega_{ext}\cap\overline{\Omega},\ \Delta<0$
 in $\Omega_{int}$  near $S$.
It follows from Proposition \ref{pro:1.1}  that the equation of $S$
can be written in the form $g_{00}(x)=0$  and $g_{00}(x)>0$  in
$\Omega_{ext}\cap\overline{\Omega},\ g_{00}<0$  in $\Omega_{int}$ near $g_{00}(x)=0$.
We shall write often the equation of $S$  in the form $S(x)=0$ and we assume that 
$\frac{\partial S(x)}{\partial x}$  is  an outward normal to the surface $S(x)=0$. 

In the case of Gordon metric the equation of $S$ has the form 
$$
(n^2(x)-1)\sum_{j=1}^n(v^j(x))^2-1=0,
$$
or, equivalently,
\begin{equation}                         \label{eq:3.2}
\sum_{j=1}^nw_j^2(x)-\frac{c^2}{n^2(x)}=0,
\end{equation}
and the domain $\Omega_{int}$  is given by the inequality
$$
\sum_{j=1}^nw_j^2(x)>\frac{c^2}{n^2(x)}
$$
near $S(x)=0$.

Suppose that $S(x)=0$  is a characteristic surface of the equation (\ref{eq:1.1}),
i.e.
\begin{equation}                                  \label{eq:3.3}
\sum_{j,k=1}^n g^{jk}(x)S_{x_j}(x)S_{x_k}(x)=0\ \ \ \mbox{when}\ \ \ S(x)=0.
\end{equation}
Let $x_j=x_j(s),\  \xi_j=\xi_j(s),\ 0\leq j\leq n,\ s\geq 0$
be a null-bicharacteristic  of (\ref{eq:1.1}),
i.e.
\begin{equation}                                \label{eq:3.4}
\frac{dx_j}{ds}=2\sum_{k=0}^ng^{jk}(x(s)\xi_j(s), \ \ x_j(0)=y_j,\ \ 0\leq j\leq n,
\end{equation}
\begin{equation}                                \label{eq:3.5}
\frac{d\xi_p}{ds}=-\sum_{j,k=0}^ng^{jk}_{x_p}(x(s))\xi_j(s)\xi_k(s)\ \ \xi_j(0)=\eta_j,
\ \ \ 0\leq p\leq n,
\end{equation}
where $x(s)=(x_1(s),...,x_n(s)),\ \eta_0=0,\ \eta\neq 0.$
The null-bicharacteristic means
that
\begin{equation}                               \label{eq:3.6}
\sum_{j,k=0}^ng^{jk}(x(s))\xi_j(s)\xi_k(s)=0.
\end{equation}
Note that if 
\begin{equation}                                \label{eq:3.7}
\sum_{j,k=0}^ng^{jk}(y)\eta_j\eta_k=0
\end{equation}
then (\ref{eq:3.6}) holds for all $s\in \R$.  

The bicharacteristic (null-bicharacteristic)  is a curve in 
$T_0(\R^{n+1})=\R^{n+1}\times(\R^{n+1}\setminus\{0\})$  and its projection on $\R^{n+1}$  is called a 
geodesic (null-geodesic).  We shall show now that the null-geodesic  satisfies the equation:
\begin{equation}                                  \label{eq:3.8}
\sum_{j,k=0}^ng_{jk}(x(s))\frac{dx_j}{ds}\frac{dx_k}{ds}=0.
\end{equation}
Denote by $G$  the matrix $[g^{jk}(x)]_{j,k=0}^n$.
Then 
$G^{-1}=[g_{jk}(x)]_{j,k=0}^n$.
It follows from (\ref{eq:3.4})  that $\frac{d\vec{x}}{ds}=2G\vec{\xi}(s)$,  where 
$\vec{x}(s)=(x_0(s),x(s)),\ \vec{\xi}=(\xi_0,\xi)$.  Then
$\vec{\xi}=\frac{1}{2}G^{-1}\frac{d\vec{x}}{ds}$  and (\ref{eq:3.6})  implies
\begin{equation}                                           \label{eq:3.9}
0=(G\vec{\xi},\vec{\xi})=
\frac{1}{4}\left(GG^{-1}\frac{d\vec{x}}{ds},G^{-1}\frac{d\vec{x}}{ds}\right)=
\frac{1}{4}
\left(G^{-1}\frac{d\vec{x}}{ds},\frac{d\vec{x}}{ds}\right),
\end{equation}
that is equavalent to (\ref{eq:3.8}).

Fix an arbitrary 
point $y$  on $S(x)=0$.  Denote by $K^+(y)$  the following half-cone in $\R^{n+1}$:
\begin{equation}                                            \label{eq:3.10}
K^+(y)=\{\vec{\xi}=(\xi_0,\xi)\in \R^{n+1}:(G(y)\vec{\xi},\vec{\xi})\geq 0,\ \xi_0\geq 0\}.
\end{equation}
Let $K_+(y)$ be the dual half-cone in $\R^{n+1}$:
\begin{equation}                                            \label{eq:3.11}
K_+(y)=\{\vec{\dot{x}}=(\dot{x}_0,\dot{x})\in \R^{n+1}:
(G^{-1}\vec{\dot{x}},\vec{\dot{x}})\geq 0,\ \dot{x}_0> 0\}.
\end{equation}
Note that 
\begin{equation}                                           \label{eq:3.12}
(\vec{\dot{x}},\vec{\xi})\geq 0 \ \ \mbox{for any}\ \ 
\vec{\dot{x}}\in K_+(y)\ \ \mbox{and any}\ \ \
\vec{\xi}\in K^+(y).
\end{equation}
Consider the null-bicharacteristic
$(\vec{x}(s),\vec{\xi}(s))$  (c.f.  (\ref{eq:3.4}),  (\ref{eq:3.5}), (\ref{eq:3.6})) 
with the following initial conditions:
\begin{equation}                                      \label{eq:3.13}
x_0(0)=y_0,\ \ x(0)=y,\ \ \xi_0(0)=0,\ \ \xi(0)=S_x(y),
\end{equation}
where $S(y)=0$.  We have (c.f. (\ref{eq:3.4})):
\begin{equation}                                       \label{eq:3.14}
\frac{dx_0(0)}{ds}=2\sum_{j=1}^ng^{0j}(y)S_{x_j}(y).
\end{equation}
\begin{pro}                                    \label{pro:3.1}
Let $(0,b(y))$  be a characteristic direction,  i.e.
$$
\sum_{j=1}^ng^{jk}(y)b_j(y)b_k(y)=0.
$$
Suppose 
$$
\sum_{j=1}^n g^{j0}(y)b_j(y)>0.
$$
Then $(0,b(y))\in K^+(y)$  and the half-cone $K_+(y)$  is contained in the half-space
$\{(\alpha_0,\alpha_1,...,\alpha_n):\sum_{j=1}^n\alpha_jb_j(y)\geq 0\}$.
\end{pro}

{\bf Proof:}  Let $\e>0$  be small.  Then $(\e,b_1(y),b_2(y),...,b_n(y))\in K^+(y)$  
since $g^{00}(y)\e^2
+2\e\sum_{j=1}^n g^{j0}(y)b_j(y)>0$  (c.f. (\ref{eq:3.10})).
For any $(\dot{x}_0,\dot{x})\in K_+(y)$  we have (c.f.  (\ref{eq:3.12})):
$$
\dot{x}_0\e+\sum_{j=1}^n\dot{x}_j b_j(y)\geq 0,
$$
i.e.  $K_+(y)$  is contained in the half-space $\e\alpha_0+
\sum_{j=1}^n\alpha_jb_j(y)\geq 0$.
Taking the limit when $\e\rightarrow 0$  we prove Proposition \ref{pro:3.1}

It follows from the hyperbolicity of (\ref{eq:1.1})  and from 
(\ref{eq:3.3})  that  the right hand side of (\ref{eq:3.14})
is not zero.   In fact,  the hyperbolicity implies that  the equation 
$\sum_{j,k=0}^ng^{jk}(y)\xi_j\xi_k=0$  has two distinct real roots 
$\xi_0^{(1)}(\xi),\xi_0^{(2)}(\xi)$  for any $\xi\neq 0$.  Taking
$\xi=S_x(y)$  we get
$g^{00}(y)\xi_0^2+2\sum_{j=1}^ng^{0i}(y)\xi_0S_{x_j}(y)=0$,  i.e.
$\xi_0^{(1)}=0,\ \xi_0^{(2)}=-2(g^{00}(y))^{-1}\sum_{j=1}^ng^{0j}(y)S_{x_j}(y)\neq 0$.
\qed

We assumed that $S_x(x)$  is the outward normal 
to the $S(x)=0$.

Since  $S(x)$ satisfies (\ref{eq:3.3})  when $S(x)=0$ 
we have that either    
\begin{equation}                                          \label{eq:3.15}
\sum_{j=1}^ng^{0j}(y)S_{x_j}(y) >0,\ \ \forall y\ \ \ \mbox{such that}\ \ S(y)=0.
\end{equation}
or
\begin{equation}                                    \label{eq:3.16}
\sum_{j=1}^n g^{0j}(y)S_{x_j}(y)<0,\ \ S(y)=0.
\end{equation}
If (\ref{eq:3.15}) holds then by Proposition \ref{pro:3.1} 
$(0,S_x(y))\in K^+(y),\ \forall y,\ S(y)=0,$
and $K_+(y)$  is contained in the half-space $(0,S_x(y))\cdot(\dot{x}_0,\dot{x})\geq 0$. 

  All  directions of $K_+(y)$ except $(1,0,...,0)$ are
pointed inside $\Omega_{ext}\times\R$.
Therefore 
the forward domain of influence of the surface 
$\{S(x)=0\}\times \R$  does not
intersect $\Omega_{int}\times\R$.  In such case 
the surface $\{S(x)=0\}\times \R$ is called the boundary of a white hole.  Note that the forward 
domain of influence of $\Omega_{int}\times\R$  does intersect $\Omega_{ext}\times\R$.

Consider now the case when  (\ref{eq:3.16}) holds.  
Then (c.f. Proposition \ref{pro:3.1}) 
the dual half-cone $K_+(y)$  is contained in the half-space 
$\vec{\dot{x}}\cdot(0,-S_x(y))\geq 0$  and  
all $\vec{\dot{x}}\in K_+(y)$  (except  $(1,0,...,0))$  are pointed inside
  $\Omega_{int}\times\R$.
Therefore the domain of influence of $\overline{\Omega}_{int}\times \R$  is contained in 
$\overline{\Omega}_{int}\times\R$.

The surface $\{S(x)=0\}\times \R$  is called the boundary of a black hole in this case.

We proved the following theorem:
\begin{theorem}                                \label{theo:3.1}
Let $S(x)=\det[g^{jk}]_{j,k=1}^n=0$  is a closed and smooth 
characteristic surface,  $g_{00}(x)>0$
in $\Omega_{ext}\cap\overline{\Omega},\ g_{00}(x)<0$  in $\Omega_{int}$ near $S(x)=0$.  
   Let $(0,S_x(x))$  be the outward normal to the surface 
$\{S(x)=0\}\times\R$.  Then $\{S(x)=0\}\times\R$  forms a white hole if 
(\ref{eq:3.15})  holds, and a black hole if (\ref{eq:3.16})  holds.
\end{theorem}

The application of Theorem \ref{theo:3.1}  to the Gordon equation yields the following
result  (c.f. [V]):

\begin{theorem}                               \label{theo:3.2}
Let $S(x)=0$  be the surface $\sum_{j=1}^nw_j^2(x)=\frac{c^2}{n^2(x)}$  and let 
$\{S(x)=0\}\times\R$  
be a characteristic surface for the Gordon equation.  
Then $\{S(x)=0\}\times\R$  forms a white hole if
$w(x)=(w_1(x),...,w_n(x))$   is pointed  inside $\Omega_{ext}$ 
when $S(x)=0$ and
$\{S(x)=0\}\times\R$  forms a black hole if $w(x)$  is pointed inside $\Omega_{int}$  
when $S(x)=0$.
\end{theorem}

{\bf Proof:}   Since 
$\{S(x)=0\}\times \R$  is a characteristic surface for the Gordon equation 
we have (c.f. (\ref{eq:2.8})):
$$
|S_x(x)|^2=\frac{(n^2-1)}{c^2\left(1-\frac{|w|^2}{c^2}\right)}
\left(\sum_{j=1}^nw_j(x)S_{x_j}(x)\right)^2.
$$
Therefore
\begin{equation}                                   \label{eq:3.17}
(c^2-|w|^2)|S_x|^2=(n^2-1)(w(x)\cdot S_x(x))^2.
\end{equation}
Since $|w|^2=\frac{c^2}{n^2(x)}$  when $S(x)=0$
we get $|S_x|^2|w|^2=(w\cdot S_x)^2$.
This last equality holds iff $S_x(x)=\alpha(x)w(x)$  where $S(x)=0,\ \alpha(x)\neq 0$.
Since $S_x(x)$  is an outward normal we have that $\alpha(x)>0$ 
if $w(x)$  pointed outwardly,  and 
$\alpha(x)<0$  if $w(x)$  is pointed inwardly.  In the case  of Gordon equation 
$g^{0j}(x)=\frac{(n^2-1)}{c\left(1-\frac{|w|^2}{c^2}\right)}
w_j(x)$  (c.f. (\ref{eq:2.10})).
We have
$\sum_{j=1}^ng^{0j}(x)S_{x_j}(x)
=\frac{(n^2-1)}{c\left(1-\frac{|w|^2}{c^2}\right)}
\alpha(x)|w|^2.$
Therefore (\ref{eq:3.15}) holds when $w(x)$ is pointed into $\Omega_{ext}$  and (\ref{eq:3.16})
holds when $w(x)$  is pointed into $\Omega_{int}$.  
In the first case we we  have a white hole and in the second case we have a black hole.
\qed

We shall study now the impact of the existence of the black or white hole on the uniqueness
of the inverse problem.

Consider the case of a white hole.  Then the domain of the dependence (i.e. the backward
domain of influence)  of any point 
$(y_0,y)\in \overline{\Omega}_{int}\times(-\infty,+\infty)$ is contained in
$\overline{\Omega}_{int}\times (-\infty,+\infty)$.
If $u(x_0,x)$  is the solution of the initial-boundary value problem 
(\ref{eq:1.1}), (\ref{eq:1.5}), (\ref{eq:1.6})
we get  that $u=0$ in $\overline{\Omega}_{int}\times(-\infty,+\infty)$   by the
uniqueness of the Cauchy problem.  Therefore the change    of the coefficients of 
(\ref{eq:1.1}) in $\Omega_{int}\times(-\infty,+\infty)$  does not change the solution of
the initial-boundary value problem  
(\ref{eq:1.1}), (\ref{eq:1.5}), (\ref{eq:1.6})
in $(\overline{\Omega}\cap\Omega_{ext})\times(-\infty,+\infty)$  and  it does not 
change the Cauchy data on
$\partial\Omega_0\times (-\infty,+\infty)$.
Therefore the boundary data on $\partial\Omega_0\times(-\infty,+\infty)$
do not able to determine the coefficients of (\ref{eq:1.1}), (modulo 
 (\ref{eq:1.10}), (\ref{eq:1.11})) in $\Omega_{int}$,  i.e.  we have a nonuniqueness.

Consider now the case of a black hole.
Then the domain of influence of each point 
$(y_0,y)\in\overline{\Omega}_{int}\times\R$  is contained in $\overline{\Omega}_{int}\times\R$
and the domain of dependence of each point $(y_0,y)\in\overline{\Omega}_{ext}\times\R$  is
contained in $\overline{\Omega}_{ext}\times\R$.

Consider the initial-boundary value problem  (\ref{eq:1.5}), (\ref{eq:1.6})
for two hyperbolic equations $L^{(i)}u^{(i)}=0$  in $\Omega\times(-\infty,+\infty)$  whose
coefficients differ in $\Omega_{int}$.  Since  domain of dependence of any
$(y_0,y)\in\overline{\Omega}_{ext}\times\R$ is contained in 
$\Omega _{ext}\times\R$  the solutions $u^{(i)},i=1,2,$  of the initial-boundary value problem
(\ref{eq:1.5}), (\ref{eq:1.6})
are equal in $\overline{\Omega}_{ext}\times(-\infty,+\infty)$.
Therefore the $\Lambda^{(1)}$  and $\Lambda^{(2)}$  are equal on 
$\partial\Omega_0\times(-\infty,+\infty)$,  i.e.  the boundary measurements are
equal despite that the coefficients of $L^{(1)}$  and $L^{(2)}$  differ in $\Omega_{int}$.  
Therefore we again have a nonuniqueness of the solution of the inverse problem.

Change the formulation of the initial-boundary value problem allowing nonzero 
initial  condition in $\Omega_{int}\times\R$  or allowing the nonzero right hand of
(\ref{eq:1.1})  with support in $\overline{\Omega}_{int}\times\R$.  Then 
$u(x_0,x)$  will be nonzero in $\Omega_{int}\times(-\infty,+\infty)$  but this 
will not effect  the solution in $\Omega_{ext}\times\R$.

\section{Black holes inside the ergosphere.}
\label{section 4}
\init
 
Let $S(x)=0,\ \Omega_{int},\ \Omega_{ext}$  be the same as in \S3.
Borrowing the terminology from the general relativity we shall call $S(x)=0$  the ergosphere
and
$\Omega_e=\Omega_{int}\cap\{\Delta(x)<0\}\cap\Omega$  is ergoregion (c.f. [V]).
We assume that $\Delta(x)< 0$  in $\overline{\Omega}_e\setminus S$.

If $S=\{x:S(x)=0\}$  is a characteristic surface then $(\Omega_{int}\cap\Omega)\times\R$  is
either a black or white hole (c.f. Theorem \ref{theo:3.1}).
Now consider the case when  $S$ is not characteristic at any point $y\in S$,  i.e. 
\begin{equation}                                              \label{eq:4.1}
\sum_{j,k=1}^ng^{jk}(y)\nu_j(y)\nu_k(y)\neq 0,\ \ \forall y\in S,
\end{equation}
where $\nu(y)$  is the outward normal to $S$.  Note that the quadratic form 
$\sum_{j,k=1}^n g^{jk}(x)\xi_j\xi_k$
has the signature $(1,-1,...,-1)$ when $x\in\Omega_e$. 
The question is whether 
there exists a black or white hole 
inside $\Omega_e\times\R$.  Consider the case $n=2$,  i.e. the case of two space dimensions
In this case there are two families of characteristic curves $S^\pm = \mbox{const}$:
\begin{equation}                                  \label{eq:4.2}
\sum_{j,k=1}^2g^{jk}(x)S_{x_j}^\pm S_{x_k}^\pm =0
\end{equation}
in a neighborhood of any point of $\overline{\Omega}_e$.
Equation (\ref{eq:4.2}) can be factored 
and we get in the region $\Omega_e^{(1)}=\{x\in\Omega_e:\ g^{22}(x)\neq 0\}$:
$$
g^{22}(x)\frac{\partial S^\pm(x)}{\partial x_2}+\left(g^{12}(x)\pm\sqrt{-\Delta(x)}\right)
\frac{\partial S^\pm}{\partial x_1}=0.
$$
Note that $\Delta(x)=g^{11}(x)g^{22}(x)-(g^{12}(x))^2<0$  in $\Omega_e$.

We shall derive an equation for the characteristic cirves $S^\pm(x)=\mbox{const}$  that 
holds  
in a neighborhood of any point
in $\overline{\Omega}_e$.  Let
$U_1$  be the set in $\overline{\Omega}_e$  where either  $g^{22}(x)\neq 0$  or 
$g^{12}(x)>0$.  Denote by $f_{U_1}^+(x)$ a nonzero vector field 
$f_{U_1}^+(x)=(g^{12}(x)+\sqrt{-\Delta(x)},g^{22}(x)),\ x\in U_1$.
Let $U_2$  be the set where either $g^{11}(x)\neq 0$  or $g^{12}(x)<0$
and denote by $f_{U_2}^+(x)$  the nonzero vector field
$f_{U_2}^+(x)=(g^{11}(x),g^{12}(x)-\sqrt{-\Delta(x)}),\ x\in U_2$.
Let $U_i^c=\overline{\Omega}_e\setminus U_i,\ i=1,2$.
Note that $U_1^c\cap U_2^c=\{x:g^{11}=g^{22}=g^{12}=0\}=\emptyset$  since the rank
of $[g^{jk}]_{j,k=1}^2$  is at least 1.  Therefore $U_1\cup U_2=\overline{\Omega}_e$.   
Let $V_i, i=1,2,$  be a small neighborhood of $U_i^c,\ i=1,2,$ 
such that $\overline{V}_1\cap \overline{V}_2=\emptyset$.  
Denote $\tilde{U}_i=U_i\setminus V_i,i=1,2$.  Then $\tilde{U}_1\cup\tilde{U}_2=\Omega_e$.
Note that 
\begin{equation}                                        \label{eq:4.3}
\frac{g^{11}(x)}{g^{12}(x)+\sqrt{-\Delta}}=
\frac{g^{12}-\sqrt{-\Delta}}{g^{22}(x)}=\lambda(x)\neq 0\ \ \mbox{in}\ \ 
\tilde{U}_1\cap\tilde{U}_2,
\end{equation}
since $g^{11}g^{22}=(g^{12}-\sqrt{-\Delta})(g^{12}+\sqrt{-\Delta})$.

Extend $\lambda(x)$ as  nonzero function from 
$\tilde{U}_1\cap \tilde{U}_2$  to $\tilde{U}_2$  
such that $\lambda(x)$  is continuous in $\tilde{U}_2$
 and
smooth 
in $\tilde{U}_2\setminus S$,  and
define
\begin{equation}                                             \label{eq:4.4}
f^+(x)=f_{U_1}^+(x)\ \ \mbox{in}\ \ \tilde{U}_1,\ \ \ \ 
f^+(x)=\lambda^{-1}(x)f_{U_2}^+(x)\ \ \mbox{in}\ \ \tilde{U}_2.
\end{equation}
Then $f^+$ is a continuous nonzero vector field in $\overline{\Omega}_e$  that is smooth
in $\Omega_e\setminus S$.
Analogously let $f_{U_3}^-(x)=(g^{12}-\sqrt{-\Delta},g^{22}), f_{U_4}^-(x)=
(g^{11},g^{12}+\sqrt{-\Delta})$,  where  $U_3$  is the set where either $g^{22}\neq 0$
or $g^{12}<0$  and $U_4$  is the set where either $g^{11}\neq 0$  or $g^{12}>0$.
Then $f_{U}^-(x)=\lambda_1(x)f_{U}^-(x)$ in 
$\tilde{U}_3\cap\tilde{U}_4$ and $\lambda_1\neq 0$  in $\tilde{U}_3\cap\tilde{U}_4$.
Here $\tilde{U}_3\subset U_3,\ \tilde{U}_4\subset U_4$  are similar to $\tilde{U}_i,i=1,2,$  in
(\ref{eq:4.4}).  Extending $\lambda_1(x)$ from $\tilde{U}_3\cap\tilde{U}_4$ to
$\tilde{U}_4$
we get a vector field $f^-(x)$ in $\overline{\Omega}_e$  that is continuous
in $\overline{\Omega}_e$ and smooth in $\overline{\Omega}_e\setminus S$.

We have that (\ref{eq:4.2}) is equivalent to
\begin{equation}                                    \label{eq:4.5}
f_1^\pm(x)S_{x_1}^\pm(x)+f_2^\pm(x)S_{x_2}^\pm(x)=0,
\end{equation}
where $f^\pm(x)=(f_1^\pm(x),f_2^\pm(x))$.  Note that  $f^\pm(x)\neq (0,0)$
for any $x\in \overline{\Omega}_e$  and $f^+(x)\neq f^-(x),\forall x\in \Omega_e\setminus S$.

Since the rank of $[g^{jk}(y)]_{j,k=1}^n$  is 1 on $S$ there exists a smooth and nonzero 
$b(y)=(b_1(y),b_2(y))$  such that
\begin{equation}                                     \label{eq:4.6}
\sum_{k=1}^2g^{jk}(y)b_k(y)=0,\ \ j=1,2,\ \ y\in S.
\end{equation}
Since
\begin{equation}                                     \label{eq:4.7}
\sum_{j,k=1}^2g^{jk}(y)b_j(y)b_k(y)=0
\end{equation}
we get from (\ref{eq:4.1}) that 
$b(y)$  is  not colinear  with $\nu(y),\ \forall y\in S$.  It follows from 
(\ref{eq:4.5})  with $S_x^\pm(y)$  replaced by $b(y)$  that
\begin{equation}                                    \label{eq:4.8}
f_1^\pm(y)b_1(y)+f_2^\pm(y)b_2(y)=0,\ \ \forall y\in S.
\end{equation}
Note that $f^+(y)=f^-(y)$  when $y\in S$.  Therefore the condition (\ref{eq:4.1})
implies  that $f^\pm(y)$  is not tangential to $S$  for any $y\in S$.  Therefore changing 
$f^\pm(x)$  to $-f^\pm(y)$  if needed we will assume that $f^\pm(y)$  is pointed 
inside $\Omega_e$ for any $y\in S$.

Consider differential equations
\begin{equation}                                    \label{eq:4.9}
\frac{d\hat{x}^\pm(\sigma)}{d\sigma}=f^\pm(x^\pm(\sigma)),\ \ \hat{x}^\pm(0)=y,\ \ y\in S,\ \ \sigma\geq 0.
\end{equation}
Let $S^\pm(x)=\mbox{const}$  be a characteristic curve.
Then (\ref{eq:4.5}), (\ref{eq:4.9})
imply that $\frac{d}{d\sigma}S^\pm(\hat{x}^\pm(\sigma))=0$,  i.e.
$S^\pm(\hat{x}^\pm(\sigma))=S^\pm(y),\sigma\geq 0$.  Therefore 
$x=\hat{x}^\pm(\sigma)$  are the parametric equations of the characteristic
curves $S^\pm(x)=S^\pm(y)$.  It was shown in \S 3  that $\sum_{j=1}^2 g^{0j}(x)b_j(x)\neq 0$
if $b(x)=(b_1,b_2)$  satisfies (\ref{eq:4.7}).  Changing $b(y)$ to $-b(y)$  if needed we assume that
\begin{equation}                                       \label{eq:4.10}
\sum_{j=1}^2g^{0j}(y)b_j(y)>0,\ \ \forall y\in S.
\end{equation}
We shall denote by $S^\pm\times\R$  the characteristic surfaces in $\Omega_e\times\R$ 
where $S^\pm(x)=\mbox{const}$ and $x_0\in \R$.  Let $(y_0,y^\pm)$  be any
point of $S^\pm\times\R$.
Consider the null-bicharacteristic 
(\ref{eq:3.4}),  (\ref{eq:3.5})
with initial conditions $x_0^\pm(0)=y_0, x^\pm(0)=y^\pm,
\xi^\pm(0)=S_x^\pm(y^\pm),\xi_0^\pm(0)=0$.
Note that $x=x^\pm(s),x_0^\pm=x_0(s),s\geq 0$  is the corresponding null-geodesic.
We shall show that $S^\pm(x^\pm(s))=S^\pm(y)$,  i.e.
 this null-geodesic  remains on the characteristic surface 
$S^\pm\times\R$.
We have
\begin{equation}                                      \label{eq:4.11}
\sum_{j,k=0}^2g^{jk}(x^\pm(s))\xi_j^\pm(s)\xi_k^\pm(s)=0,
\end{equation}
since 
$\sum_{j,k=0}^2g^{jk}\eta_j^\pm\eta_k^\pm=0,\ \eta_0^\pm=0,\ \eta^\pm=S_x^\pm(y^\pm)$ 
(c.f. (\ref{eq:3.6}),
(\ref{eq:3.7})).
Note that $\xi_0^\pm(s)=\xi^\pm(0)=0$  since $[g^{jk}(x)]_{j,k=0}^2$
is independent of $x_0$ (c.f. (\ref{eq:3.5})).
Since $S^\pm(x)$  is a solution of  
the eiconal equation 
$\sum_{j,k=1}^2g^{jk}(x)S_{x_j}^\pm(x)S_{x_k}^\pm(x)=0$ in an open neighborhood of
$y^\pm$  in $\Omega_e$,  we have that (c.f. [CH])
$$
\xi^\pm(s)=S_x^\pm(x^\pm(s)).
$$ 
We have 
$\left(\frac{dx_0^\pm}{ds},\frac{dx^\pm}{ds}\right)=2G(0,\xi^\pm(s))$.
Therefore 
$$
\sum_{j=1}^2\frac{dx_j^\pm}{ds}S_{x_j}^\pm(x^\pm(s))=
\sum_{j=1}^2\frac{dx_j^\pm}{ds}\xi_j^\pm(s)=2\sum_{j,k=0}^2g^{jk}(x^\pm(s))\xi_j^\pm(s)\xi_k(s)=0,
$$
i.e.  $\frac{d}{ds}S^\pm(x^\pm(s))=0,\ \forall s.$

Therefore the projection of the null-bicharacteristic $x_0=x_0^\pm(s), x=x^\pm(s),x^\pm(0)=y^\pm,
\xi_0=0,\xi=\xi^\pm(s)$ on the 
$(x_1,x_2)$-plane   belongs to the curve $S^\pm(x)=S^\pm(y^\pm)$.

Fix arbitrary $y\in S$.  Denote by $\Pi^+(y)$
the intersection of the half-plane
\begin{equation}                               \label{eq:4.12}
\xi_1b_1(y)+\xi_2b_2(y)>0
\end{equation}
with $\Omega_e$.   Analogously let $\Pi^-(y)$  be the intersection of the half-plane
\begin{equation}                                \label{eq:4.13}
\xi_1b_1(y)+\xi_2b_2(y)<0
\end{equation}
with $\Omega_e$.
Let $x=\hat{x}^\pm(\sigma)$  be the solution of (\ref{eq:4.9}), $\hat{x}^\pm(0)=y,\sigma\geq 0$.
Note that $f^+(y)=f^-(y)\in\overline{\Pi}^+(y)\cap\overline{\Pi}^-(y)$.
However for $\sigma >0$  and small, either $\hat{x}^+(\sigma)$    belongs to
$\Pi^+(y)$ and $\hat{x}^-(\sigma)$     to $\Pi^-(y)$,  or vice versa.  For the definiteness let
$\hat{x}^+(\sigma)\in\Pi^+(y)$  for $0<\sigma<\e$ and $\hat{x}^-(\sigma)$  belongs   
to $\Pi^-(y)$   for $0<\sigma<\e$.
Condition (\ref{eq:4.1}) and the continuity in $y$ imply 
$\hat{x}^+(\sigma)\in \Pi^+(y),\ \hat{x}^-(\sigma)\in \Pi^-(y),\ 0<\sigma<\e$,
for all $y\in S$.
Each $\hat{x}^\pm(\sigma)$  is a parametric equation of the characteristic curve 
$S^\pm(x)=S^\pm(y)$,  i.e. 
\begin{equation}                                 \label{eq:4.14}
S^\pm(\hat{x}^\pm(\sigma))=S^\pm(y),\ \ \forall\sigma\geq 0.
\end{equation}
Note that 
 the curve $S^\pm(x)=S^\pm(y)$ is contained in $\Pi^\pm(y)$
when $x\neq y$  and close to $y$.

Consider now two null-bicharacteristics
\begin{equation}                                 \label{eq:4.15}
x_0=x_0^+(s),\ \ x=x^+(s),\ \ \xi_0=0,\ \ \xi=\xi^+(s),\ s\geq 0,
\end{equation}
with initial conditions $x_0^+(0)=0,x^+(0)=y,\xi^+(0)= b$
and
\begin{equation} 
\nonumber                                
(4.15')\ \ \ \ \ \ \ \ \ 
x_0=x_0^-(s),\ \ x=x^-(s),\ \ \xi_0=0,\ \ \xi=\xi^-(s),\ s\geq 0,
\ \ \ \ \ \ \ \ \ \ \ 
\end{equation}
with initial conditions $x_0^-(0)=0,x^-(0)=y,\xi^-(0)=- b$.

Note that 
\begin{equation}                                  \label{eq:4.16}
\frac{dx_0^+(s)}{ds}=2\sum_{j=1}^2g^{j0}(x^+(s))\xi^+_j(s)>0
\end{equation}
since (\ref{eq:4.10}) holds.  Also we have that
\begin{equation}                                 \label{eq:4.17}
\frac{dx_0^-(s)}{ds}<0,\ \ s\geq 0,
\end{equation}
since $\xi^-(0)=-b$.  Therefore $x_0=x_0^+(s),\ x=x^+(s)$ is a forward null-geodesics
since $x_0$   is increasing when $s$  is increasing and $x=x_0^-(s),\ x=x^-(s)$
is a backward null-geodesics.

It follows from (\ref{eq:3.5}) and (\ref{eq:4.6})
that 
$$
\frac{dx_j^\pm(0)}{ds}=2\sum_{k=1}^2g^{jk}(y)b_k(y)=0,\ j=1,2.
$$
However
$$
\frac{dx^\pm(s)}{ds}=2G(x^\pm(s))(0,\xi^\pm(s))\neq 0,\ \ s>0,
$$
since $\Delta(x)\neq 0$  in $\Omega_e\setminus S$.

Since $(0,b(y))\in K^+(y)$  (c.f. (\ref{eq:3.10}))  the dual half-cone $K_+(y)$  is
contained in $\overline{\Pi}^+(y)$.  Therefore 
 the projection $x=x^+(s),\ s\geq 0,$  of the null-bicharacteristic 
(\ref{eq:4.15}) satisfies
\begin{equation}                                        \label{eq:4.18}
S^+(x^+(s))=S^+(y),\ s\geq 0.
\end{equation}
Analogously,  the projection $x=x^-(s)$  of the null-bicharacteristic (4.15$'$) on $(x_1,x_2)$-plane
satisfies
\begin{equation}
\nonumber                                        
(4.18')\ \ \ \ \ \ \ \ \ \ \ \ \ \ \ \ \ \ \ \ \ \ \ \ S^-(x^-(s))=S^-(y),\ s\geq 0.
\ \ \ \ \ \ \ \ \ \ \ \ \ \ \ \
\end{equation}

Comparing (\ref{eq:4.14}) and (\ref{eq:4.18}), (4.18$'$) we get that 
$x=x^\pm(s), x=\hat{x}^\pm(\sigma)$  are
different parametrization of the same curve,  i.e.
\begin{equation}                                       \label{eq:4.19}
x^\pm(s^\pm(\sigma))=\hat{x}^\pm(\sigma),\ \ \sigma\geq 0,
\end{equation}
where  $s^\pm(0)=0,\ \frac{ds^\pm(\sigma)}{d\sigma}>0$  for $\sigma>0$.
\qed  

{\bf Remark 4.1} Note that if $x=x_0(s),x=x(s),\xi_0=0,\xi=\xi(s),$  satisfies 
(\ref{eq:3.4}), (\ref{eq:3.5}),  then $x=x_0(-s),x=x(-s),\xi_0=0,\xi=-\xi(-s)$ also
satisfies (\ref{eq:3.4}), (\ref{eq:3.5}).
Therefore changing $s\geq 0$  to $-s$  in (4.15$'$) and 
combining (\ref{eq:4.15}) and (4.15$'$)  we get a forward null-bicharacteristic
$x=x_0(s),x=x(s),\xi_0=0,\xi=\xi(s),$  defined on $(-\infty,+\infty)$  with initial
conditions  $x_0(0)=0,x(0)=y,\xi(0)=b$.  The projection  of this 
bicharacteristic on $(x_1,x_2)$-plane  has a singularity (caustic)  at $x=y$. 
\qed

Let $\Omega_e\subset \Omega\cap\Omega_{int}\cap\{\Delta<0\}$.
We assume that the boundary of $\Omega_e$ consists of $S=\partial\Omega_{int}$  and
$S_1$.  Assume that $\Delta<0$  on $S_1$ and $\Omega_e$  is diffeomorphic to an annulus domain in
$\R^2$.

Let 
$y$ be an arbitrary point on $S_1$.  Consider the forward cone
of influence $K_+(y)$. Let $K_+^{(0)}(y)$ be the projection of this cone 
on the $(x_1,x_2)$-plane.   We assume that
\begin{equation}                                 \label{eq:4.20}
N(y)\cdot\dot{x}>0,\ \ \forall y\in S_1, \ \forall \dot{x}\in K_+^{(0)}(y),
\end{equation}
where $N(y)$ is the outward unit normal to $S_1$  and $\dot{x}=(\dot{x}_1,\dot{x}_2)\in\R^2$
is any vector in $K_+^{(0)}(y)$.  In particular,  $S_1$  is not a characteristic curve.
Note that condition (\ref{eq:4.20}) is equivalent to the condition that
\begin{equation}                                 \label{eq:4.21}
N(y)\cdot\frac{dx^\pm(s_1^\pm)}{ds}>0,\ \ x^\pm(s_1^\pm)=y,\ \forall y\in S_1,
\end{equation}
where $x^\pm(s)$  is the projection on $(x_1,x_2)$-plane  of two forward null-bicharacteristics
such that $x=x^\pm(s)$  are parametric equations of two characteristics 
of the form (\ref{eq:4.2})
passing through
$y\in S_1$.   Note that $\frac{dx_0^\pm(s_1^\pm)}{ds}>0$ when $x^\pm(s_1^\pm)\in S_1$.

We shall say that conditions (4.20') or (4.21')  are satisfied if $N(y)\cdot\dot{x}<0$  in 
(\ref{eq:4.20})  or $N(y)\cdot\frac{dx^\pm(s)}{ds}<0$  hold in (\ref{eq:4.21}).
\begin{theorem}                                         \label{theo:4.1}
Suppose $\partial\Omega_e=S_0\cup S_1$,  where $\Delta(x)=0$  on $S,\ \Delta(x)<0$  on 
$\overline{\Omega}_e\setminus S$.  Suppose (\ref{eq:4.1})  holds on $S$.
Suppose also that either (\ref{eq:4.20}) or (4.20$'$)  is satisfied  on $S_1$.  Then there
exists a Jordan curve $S_0(x)=0$  between $S$ and $S_1$  such that 
$S_0\times \R$  is a characteristic surface,  i.e. $S_0\times \R$  is a boundary of either a black or
a white hole.
\end{theorem}

{\bf Proof:} It was shown already that condition (\ref{eq:4.1}) implies the existence of 
two null-bicharacteristics
$x_0=x_0^\pm(s),x=x^\pm(s),\xi_0^\pm=0,\xi=\xi^\pm(s),s\geq 0$,  such that $x^\pm(0)=y\in S$
and $x^\pm(s)$  after reparametrization $s=s^\pm(\sigma),s^\pm(0)=0,\frac{ds^\pm(\sigma)}{d\sigma}>0,
\sigma>0$,  coincide with the solution 
$x=\hat{x}^\pm(\sigma)$
  of the differential 
equation (\ref{eq:4.9}):
$x^\pm(s(\sigma))=\hat{x}^\pm(\sigma),\sigma\geq 0$.
Moreover $\frac{dx_0^-(s)}{ds}<0$,  i.e.  $x_0$  decreases when $\sigma$ (or $s$)  increases,
and $\frac{dx_0^+(s)}{ds}>0$.

Suppose condition (\ref{eq:4.20}) is satisfied.
We shall show that there is no solution $x=\hat{x}^-(\sigma)$ of
\begin{equation}                                       \label{eq:4.22}
\frac{d\hat{x}^-(\sigma)}{d\sigma}=f^-(\hat{x}(\sigma)),
\end{equation}
$\hat{x}(0)=y\in S$  that reaches $S_1$,  i.e.  $\hat{x}^-(\sigma_1)=y^{(1)}$  where 
$y^{(1)}\in S_1,\sigma_1>0.$

Suppose such $x=\hat{x}^-(\sigma)$ exists.  When $\sigma>\sigma_1$   $\ \ x=\hat{x}^-(\sigma)$
leaves $\Omega_e$  since $S_1$  is not characteristic.  
Note that for the null-bicharacteristic whose
projection is $x=x^-(s(\sigma))=\hat{x}(\sigma)$,  the time variable $x_0=x_0^-(s)$ is
decreasing when $\sigma$  is increasing.  Therefore $x=\hat{x}^-(\sigma)$  leaves $\Omega_e$
when $x_0$  is decreasing.  From other side the condition (\ref{eq:4.20})  implies that
the projection 
of all null bi-characteristics passing through $y^{(1)}\in S_1$  leave $\Omega_e$  when $x_0$  increases.
This contradiction proves that the limit set of the trajectory 
$x=\hat{x}^-(\sigma)$ is contained inside $\Omega_e$.
By the Poincare-Bendixson theorem  (c.f. [H]) there exists a limit cycle,  i.e. a closed periodic solution
$x=z^-(\sigma)$   of (\ref{eq:4.22})  
which has no points of self-intersection.  Let $S_0(x)=0$  be the equation of this orbit.  Then
$S_0\times \R$  be a characteristic surface where $S_0=\{x:S_0(x)=0\}$.  Other solutios 
of (\ref{eq:4.22})  are spiraled around $S_0$  when $x_0\rightarrow\-infty$.

Now we shall assume that the condition (4.20$'$)  (or (4.21$'$)) 
is satisfied.
These conditions mean that the projection on $(x_1,x_2)$-plane of any null-bicharacteristic 
passing through $S_1$ enters $\Omega_e$  when $x_0$  increases.  From other side, consider
the solution of
\begin{equation}                                            \label{eq:4.23}
\frac{d\hat{x}^+(\sigma)}{d\sigma}=f^+(\hat{x}^+(\sigma)),\ \ \sigma\geq 0,
\end{equation}
that starts on $S:\hat{x}(0)=y\in S$.  We claim that this solution 
can not reach $S_1$.  In fact if $x=\hat{x}^+(\sigma)$ reaches $S_1$ it
will leave $S_1$  when $\sigma >\sigma_1$.  We have $\frac{dx_0^+(0)}{ds}>0$  for
the null-bicharacteristics whose projection on the $(x_1,x_2)$-plane is $x=\hat{x}^+(\sigma)$ after 
the change of the parameter $s=s^+(\sigma)$.  Therefore $x=x^+(s)$  leaves $\Omega_e$ when $x_0$ 
increases.  This contradiction shows that the limit set of $x=x^+(\sigma)$  is contained 
inside $\Omega_e$.

Again by the Poincare-Bendixson  theorem there exists a closed characteristic curve 
$S_0(x)=0$  without points of self-intersection.  All other solutions of (\ref{eq:4.22})
spiral around $S_0$
  when $x_0\rightarrow +\infty$.

{\bf Remark 4.2}
Without additional assumptions it is impossible to specify whether 
$S_0\times\R$ is a boundary of a white or a black hole.

For example,  assume that the condition (\ref{eq:4.20})  is satisfied.  Then  $S_0$  is a limit 
cycle for the 
equation $\frac{d\hat{x}^-(\sigma)}{d\sigma}=f^-(\hat{x}^-(\sigma))$.  Since $f^-(x)\neq f^+(x)$
for all $x\in \Omega_e$ we have that $f^+(x),x\in S_0$  is pointed either into
the interior of $S_0$  for all $x\in S_0$  or into the exterior of $S_0$.
In the first case $S_0\times \R$  forms a black hole and in the second case $S_0\times \R$ forms
a white hole.  In the second case we can apply the Poincare-Bendixson theorem 
to the equation (\ref{eq:4.23}) in 
 the annulus
domain between $S$  and $S_0$.  Then there exists a limit cycle $S_2$ for the
equation $\frac{d\hat{x}^+(\sigma)}{d\sigma}=f^+(\hat{x}(\sigma))$ between $S$  and $S_1$  and 
$S_2\times \R$  is the boundary of a white hole.  Also there exists a white or black hole 
formed by $S_3\times \R$  where 
$S_3(x)=0$ is a limit cycle for (\ref{eq:4.23}) 
between $S_0$  and $S_1$.  Analoguously  if condition (4.20$'$) is satisfied 
then the limit cycle $S_0$  for the equation (\ref{eq:4.23}) forms
a white hole if $f^-(x)$ restricted to $S_0$  is pointed in the exterior of $S_0$  and there
are two additional black or white holes if  $f^-(x)|_{S_0}$  is pointed into the interior of
$S_0$. 

{\bf Remark 4.3}
The black or white holes described in Theorem \ref{theo:4.1}  are stable in the sense that if
we perturb slightly $[g^{jk}(x)]_{j,k=0}^2$  then assumptions of Theorem \ref{theo:4.1}   
 remain valid and therefore the limit cycle solution will still exist.
This is in contrast with black or white holes obtained by Theorem \ref{theo:3.1}: 
when we perturb
$[g^{jk}]_{j,k=0}^3$  then the surface $S=\{\Delta(x)=0\}$ may cease to be  characteristic  
and  the black or white hole at $S\times \R$ disappears.

\qed 

Now we shall formulate the application of Theorem \ref{theo:4.1} to the case of the Gordon equation.

\begin{theorem}                                        \label{theo:4.2}
Let $\partial\Omega_e=S\cup S_1$.  Assume that 
$|w(x)|^2=\frac{c^2}{n^2(x)}$  on $S$  and $|w(x)|^2>\frac{c^2}{n^2(x)}$
in $\overline{\Omega}_e\setminus S$.  Suppose $w(y)$  is not colinear with 
$\nu(y)$  on $S$,  where $\nu(y)$  is the outward unit normal to $S$.  
Suppose that either 
\begin{equation}                                        \label{eq:4.24}
(n^2(x)-1)^{\frac{1}{2}}(v(x)\cdot N)>1\ \ \mbox{on}\ S_1,
\end{equation}
or
$$
%\begin{equation}                                        
%\label{eq:4.25}
%\nonumber
(4.24')\ \ \ \ \ \ \ \ \ \ \ \ \ \ \ \ \ \ \ \ \ \ 
(n^2(x)-1)^{\frac{1}{2}}(v(x)\cdot N)<-1\ \ \mbox{on}\ S_1,
\ \ \ \ \ \ \ \ \ \ \ \ \ \ \ \ \ \ \
$$
%\end{equation}
where $v(x)=\left(1-\frac{|w(x)|^2}{c^2}\right)^{-\frac{1}{2}}\frac{w(x)}{c},\ N(x)$  is
the outward unit normal to $S_1$.

Then there exists a Jordan curve $S_0(x)=0$  between $S$  and $S_1$  such that
$S_0\times\R$  is a characteristic surface,  i.e.  $S_0\times\R$  is 
the boundary of
either black
or white hole.
\end{theorem}

{\bf Proof:}  Note that $\Delta(y)=0$  is equivalent to $|w|^2=\frac{c^2}{n^2(y)},\ y\in S$ 
 in the case of the 
Gordon equation and $\Delta(x)<0$  implies $|w|^2>\frac{c^2}{n^2(y)}$  in 
$\overline{\Omega}_e\setminus S$.

Since $\Delta(y)=0$  there exists a smooth $b(y)\neq 0,\ y\in S$,  such 
that $\sum_{k=1}^2g^{jk}(y)b_k(y)=0,\ j=1,2 $(c.f. (\ref{eq:4.6})).
For the Gordon equation we have
$$
-b_j(y)+(n^2-1)(v\cdot b)v_j=0,\ \ j=1,2,
$$
i.e. $b(y)$  is colinear with $w(y)$  on $S$.  Therefore $b(y)$  is not colinear with $\nu(y)$  since
$w(y)$  is not colinear,  $\forall y\in S$.  Note that condition (\ref{eq:4.10}) has the form
$(n^2-1)(v\cdot b)v^0>0$,  i.e.  $b$  and $w$  has the same direction.  We take $b=w$.
Therefore $f^+(y)=f^-(y)$ is orthogonal to $w(y)$ and is not tangential to $S,\ \forall y\in S$.

Let $y^{(1)}$  be any point on $S_1$.  Suppose (\ref{eq:4.24}) holds.
Consider any null-bicharacteristic whose projection is passing through $y^{(1)}.$
The equation 
$$
\sum_{j,k=1}^2g^{jk}(y^{(1)})\xi_j(y^{(1)})\xi_k(y^{(1)})=0
$$  
has the following form 
for the Gordon equation
\begin{equation}                                  \label{eq:4.25}
-|\xi(y^{(1)})|^2+(n^2(y^{(1)})-1)(v(y^{(1)})\cdot \xi(y^{(1})))^2=0.
\end{equation} 
Note that 
$\frac{dx_0(s)}{ds}=2(n^2-1)v^0(v\cdot\xi)>0$  implies that
$(v(x(s))\cdot\xi(s))>0$.  It follows from (\ref{eq:4.25}) that
\begin{equation}                                  \label{eq:4.26}
v\cdot\xi=\frac{|\xi|}{(n^2-1)^{\frac{1}{2}}}.
\end{equation}
We have 
\begin{equation}                                \label{eq:4.27}
\ \ \ \ \ \frac{dx_j}{ds}=2\sum_{k=1}^2g^{jk}\xi_k=-2\xi_j+2(n^2-1)(v\cdot\xi)v_j=-2\xi_j+
2(n^2-1)^{\frac{1}{2}}|\xi|v_j,\ \ j=1,2.
\end{equation}
Take the inner product of
(\ref{eq:4.27}) with $N$,  where $N$  is outward unit normal to $S_1$.
We get
$$
\frac{dx}{ds}\cdot N=-2\xi\cdot N+2|\xi|(n^2-1)^{\frac{1}{2}}(v\cdot N).
$$
Since  $|\xi\cdot N|\leq |\xi|$  we get,  using (\ref{eq:4.24}),  that 
$\frac{dx}{ds}\cdot N>0$  on $S_1$.  Note that   $\frac{dx_0(s)}{ds}>0$.
Therefore any null-bicharacteristic escapes $\Omega_e$  when $s>s_0$  (or when $x_0>x_0(s_0)$)
where $x(s_0)\in S_1$.

Therefore by Theorem \ref{theo:4.1}  we have a characteristic surface $S_0\times\R$.  If
$(n^2-1)^{\frac{1}{2}}(v\cdot N)<-1$  then all null-bicharacteristics are entering $\Omega_e$
when $x_0$  increases.  Therefore again we can apply Theorem \ref{theo:4.1}.

As  in Remark 4.3 the black and white holes described in Theorem \ref{theo:4.2}
are stable.

Now we shall refine Theorem  \ref{theo:4.2} and specify when 
the characteristic surface   
$S_0\times\R$  is the boundary of a white hole and when it is the boundary of a black hole.

We shall say that the flow $w(x)=(w,(x),w_2(x))$  is incoming if for any closed 
simple curve $\Gamma\subset\Omega_e$  containing $S_1$  there exists at least one point 
$y\in \Gamma$  such that $w(y)\cdot \nu(y)>0$  where $\nu(y)$  is a normal to $\Gamma$
pointed inside $\Gamma$.   Analogously,  the flow $w(x)$ is called outgoing in $\Omega_e$  if 
for any $\Gamma$  there exists $y\in \Gamma$  such that $w(y)\cdot \nu(y)<0$.

\begin{theorem}                                 \label{theo:4.3}
Consider the Gordon equation in $\Omega_e$ where $\Omega_e$  is the same as in Theorems
\ref{theo:4.1} and \ref{theo:4.2}.  Suppose 
$w(y)$ is not colinear with the normal to $S,\ \forall y\in S$,
and (\ref{eq:4.24}) holds on $S_1$.  Assume in addition that
the flow $w(x)$  is incoming. Then there exists a black hole  
bounded by
$S_0\times\R$ where
$S_0$ is a Jordan curve between $S$  and $S_1$.  If  (4.24')  holds and 
the flow $w(x)$ is outgoing then there exists a white hole with the boundary $S_0\times \R$.
\end{theorem}

{\bf Proof:}  Suppose (\ref{eq:4.24})  holds.  It was proven in Theorem \ref{theo:4.2} that
there exists a characteristic surface $S_0\times\R$.  Now using that the flow $w(x)$  is incoming 
we shall prove that $S_0\times\R$  is the boundary of a black hole.

Take any $y\in S_0$. Let $b_1(y)$  be a normal to $S_0$.  Choose the direction of $b_1(y)$  such that
$\sum_{j=1}^2g^{0j}(y)b_j(y)=
2(n^2(y)-1)v^0(y)(v(y)\cdot b_1(y))>0$,  i.e.  
$b_1(y)\cdot w(y)>0)$.  It follows from the Proposition \ref{pro:3.1}
that $K_+(y)$ (c.f. (\ref{eq:3.11}))
is contained in the half-space $(\alpha_0,\alpha_1,\alpha_2)\cdot (0,b_1(y),b_2(y))\geq 0$.

Since 
the flow $w(x)$ is incoming there exists $y_0\in S_0$  such that $w(y_0)\cdot \nu(y_0)>0$ 
where $\nu(y_0)$  is pointed inside $S_0$.  Since $b_1(y_0)\cdot w(y_0)>0$ 
and $b_1(y_0)$  is colinear with $\nu(y_0)$  we get that $b_1(y_0)$ is also pointed
inside $S_0$.  Since $b_1(y)$ is continuous in $y$  we have that $b_1(y)$ is pointed
inside $S_0$ for all $y\in S_0$.  Therefore  $(\dot{x}_0,\dot{x})\in K_+(y)$  
are pointed inside $S_0\times \R$,  i.e.  $S_0\times\R$  is the boundary of a black hole.

Suppose that (4.24') is satisfied and the flow $w(x)$  is outgoing.  Let
$b_1(y)$  be the same as above,  i.e.  $b_1(y)$ is a normal to $S_0$ and $b_1(y)\cdot w(y)>0$.
Since the flow $w(x)$  is outgoing there exists $y_0\in S_0$  such that 
$w(y_0)\cdot \nu(y_0)<0$  where
$\nu(y_0)$ is pointed inside $S_0$.  Then $b_1(y_0)$  is a normal to $S_0$ that is pointed 
outside of $S_0$.
Therefore $K_+(y_0)$  is pointed outside of $S_0\times\R$.
The same is true for any $y\in S_0$  by the continuity.
Therefore $S_0\times \R$  is the boundary of a white hole.
\qed

{\bf Example 4.1 }(Acoustic black hole (c.f. [V])).

Consider a fluid flow in a vortex with the velocity field
\begin{equation}                                      \label{eq:4.28}            
v=(v^1,v^2)=\frac{A}{r}\hat{r}+\frac{B}{r}\hat{\theta},
\end{equation} 
where $r=|x|,\ \hat{r}=\left(\frac{x_1}{|x|},\frac{x_2}{|x_2}\right),\ 
\theta=\left(-\frac{x_2}{|x|},\frac{x_1}{|x|}\right),\ A$ and $B$ are constants.
The inverse of the metric tensor in this case has the form
\begin{eqnarray}                                  \label{eq:4.29}
g^{00}=\frac{1}{\rho c},\ \ 
g^{0j}=g^{j0}=\frac{1}{\rho c}v^j,\ 
1\leq j\leq 2,
\\
\nonumber
g^{jk}=\frac{1}{\rho c}(-c^2\delta_{ij}+v^iv^j),\ \ 1\leq j,k \leq 2,
\end{eqnarray}
where $c$  is the sound speed,  $\rho$  is the density.

It is  
convenient to use polar coordinates in (\ref{eq:4.2}).  Assuming $c=1,\rho=1$  we have
\begin{equation}                                    \label{eq:4.30}
\left(\frac{A^2}{r^2}-1\right)\left(\frac{\partial S}{\partial r}\right)^2+
\frac{2AB}{r^2}\ \frac{\partial S}{\partial r}\ \frac{\partial S}{\partial \theta}
+\left(\frac{B^2}{r^4}-\frac{1}{r^2}\right)\left(\frac{\partial S}{\partial \theta}\right)^2=0,
\end{equation}
or (c.f. (\ref{eq:4.5}))
\begin{equation}                                    \label{eq:4.31}
\left(\frac{A^2}{r^2}-1\right)\frac{\partial S^\pm}{\partial r}+
\left(\frac{AB}{r^3}\pm\sqrt{\frac{A^2+B^2}{ r^4}-\frac{1}{r^2}}
     \right)\frac{\partial S^\pm}{\partial \theta}
\end{equation}
for $r<\sqrt{A^2+B^2}$.
The ergosphere,
i.e. the curve where $\Delta=0$  has the form 
$A^2+B^2=r^2$.
We shall consider the domain $0<r<\sqrt{A^2+B^2}$  or any compact domain $\overline{\Omega}_e=
\{r_1\leq r\leq r_0\}$  where $r_0=\sqrt{A^2+B^2},\ r_1<|A|$.  The condition (\ref{eq:4.1}) is
satisfied when $B\neq 0$.  
The  condition (\ref{eq:4.20})  is satisfied when $A>r_1$  and the condition (4.20') 
holds when $A<-r_1$.

Consider the case $A>0,B>0$.  The system  (\ref{eq:4.9})  has the following form
in polar coordinates:
\begin{equation}                                         \label{eq:4.32}
\frac{dr^+}{ds}=A^2-r^2, \ \ \frac{d\theta^+}{ds}=\frac{AB}{r}+\sqrt{A^2+B^2-r^2},
\end{equation}
\begin{equation}                                          \label{eq:4.33}
\frac{dr^-(s)}{ds}=-1,\ \ \frac{d\theta^-(s)}{ds}=
\frac{1-\frac{B^2}{r^2}}{\frac{AB}{r}+\sqrt{A^2+B^2-r^2}},
\end{equation}
$r^\pm(0)=r_0,\ \theta^\pm(0)=\frac{y}{|y|},\ |y|=r_0,\ s\geq 0$.
Note that $\frac{dx}{ds}=\frac{dr}{ds}\hat{r}+r\frac{d\theta}{ds}\hat{\theta}$.
We wanted to ensure that the vector field in the right hand sides of (\ref{eq:4.32}),
(\ref{eq:4.33})
is not zero.  Since $\frac{AB}{r}-\sqrt{A^2+B^2-r^2}=0$  when $A=r,B>0$  we divided by
$r^2-A^2$  in (\ref{eq:4.33})  and used that 
$\frac{AB}{r}-\sqrt{A^2+B^2-r^2}
=\frac{(r^2-A^2)\left(1-\frac{B^2}{r^2}\right)}{\frac{AB}{r}+\sqrt{A^2+B^2-r^2}}$. 
It is clear that $r^2=A^2$  is the limit cycle for the equation (\ref{eq:4.32}).

In the notations of Theorem \ref{theo:4.2} (c.f. (\ref{eq:4.10}),
(\ref{eq:4.29}))   we have $b(y)=v(y)=\frac{A}{r_0}\hat{r}+
\frac{B}{r_0}\hat{\theta}$  where $|y|=r_0=\sqrt{A^2+B^2}$.  We have,   (c.f.  (\ref{eq:4.12}),
(\ref{eq:4.13})):
$$\frac{dx^+}{ds}\cdot b(y)=\left(\frac{dr^+}{ds}\hat{r}+
r_0\frac{d\hat{\theta}(s)}{ds}\hat{\theta}\right)\cdot
\left( \frac{A}{r_0}\hat{r}+\frac{B}{r_0}\hat{\theta}\right)
$$
$$
=-\frac{AB^2}{r_0}+\frac{AB^2}{r_0}+B\sqrt{r_0^2-r^2(s)}+O(r_0-r(s))>0\ \ \mbox{for}\ \ 
0\leq s\leq \e.
$$
Analogously $\frac{dx^-(s)}{ds}\cdot b(y)< 0$  for $0< s<\e,\ \e$  is small.
Note that $r=r^+(s),\theta=\theta^+(s)$  is the projection on $(x_1,x_2)$-plane  (after 
a reparametrization)
of the null-bicharacteristic with initial conditions 
$x^+(0)=y,|y|=r_0,x_0^+(0)=0,\xi_0^+(0)=0,\xi(0)=b(y)$
and $\frac{dx_0^+(s)}{ds}>0,\ \forall s\geq 0$.

When $x_0\rightarrow +\infty,\ \ r=r^+(s),\theta=\theta^+(s)$  spirals toward 
the limit cycle $r=A$.  Also $r=r^-(s),\theta=\theta^-(s)$  is the projection (after
a reparametrization)  of the null-bicharacteristic with initial conditions 
$x_0^-(0)=0,x^-(0)=y,\xi_0^-=0,\xi(0)=-b(y)$  and  $\frac{dx_0^-(s)}{ds}<0$.
Note that $r=r^-(s),\theta=\theta^-(s)$  reaches $r=r_1$  when $x_0\rightarrow -\infty$.
Therefore $r=A$  is the boundary of a white hole.

When $A<-r_1, \ B>0$  we have,  analogously to (\ref{eq:4.32}), (\ref{eq:4.33}):
\begin{equation}                                         \label{eq:4.34}
\frac{dr^+}{ds}=-1, \ \ \frac{d\theta^+}{ds}=
\frac{1-\frac{B^2}{r^2}}{\frac{AB}{r}-\sqrt{A^2+B^2-r^2}},
\end{equation}
\begin{equation}                                          \label{eq:4.35}
\frac{dr^-(s)}{ds}=A^2-r^2,\ \ \frac{d\theta^-(s)}{ds}=
\frac{AB}{r}-\sqrt{A^2+B^2-r^2},
\end{equation}
$r^\pm(0)=r_0,\ \theta^\pm(0)=\frac{y}{|y|},\ |y|=r_0$,
and we modified (\ref{eq:4.34}) since $\frac{AB}{r}+\sqrt{A^2+B^2-r^2}=0$
when $r=|A|$.  Now $r=r^-(s),\ \theta=\theta^-(s)$  spirals towards  $r=|A|$  
when $x_0\rightarrow -\infty$
and  $r=r^+(s),\theta=\theta^+(s)$ reaches $r=r_1$ when $x_0\rightarrow +\infty$.
Therefore $r=|A|$  is the boundary of a black hole.

A similar results hold when $B<0$.
The difference is that the spiraling
of the solutions towards $r=|A|$  changes from the clockwise to 
the counter-clockwise
or vice versa.

{\bf Example 4.2}  Consider a generalization of Example 4.1 
when
$$
v(x)=A(r)\hat{r}+B(r)\hat{\theta},
$$
where
$r_1\leq r\leq r_0,\ A(r), B(r)$  are smooth,  $A^2(r_0)+B^2(r_0)=1,\ A^2(r)+B^2(r)>1$
on $[r_1,r_0),\ B(r)>0$  on $[r_1,r_0],\ A(r)+1$  has simple zeros $\alpha_1,...,\alpha_{m_1}$
on $(r_1,r_0),\ A(r)-1$  has simple zeros $\beta_1,...,\beta_{m_2}$ on $(r_1,r_0),\beta_k\neq\alpha_j,
\forall j,k,\ |A(r_1)|>1$.  Equation (\ref{eq:4.31})  has the form
\begin{equation}                                    \label{eq:4.36}
(A^2(r)-1)S_r^\pm + (A(r)B(r)\pm\sqrt{A^2+B^2-1})S_\theta^\pm=0.
\end{equation}  
Let
\begin{equation}                                   \label{eq:4.37}
\frac{dr^+(s)}{ds}=A(r)-1,\ \ 
\frac{d\theta^+}{ds}=\frac{A(r)B(r)+\sqrt{A^2+B^2-1}}{A(r)+1},
\end{equation}
\begin{equation}                                   \label{eq:4.38}
\frac{dr^-(s)}{ds}=A(r)+1,\ \ 
\frac{d\theta^-}{ds}=\frac{A(r)B(r)-\sqrt{A^2+B^2-1}}{A(r)-1}.
\end{equation}
Note that the right hand sides in (\ref{eq:4.37}), (\ref{eq:4.38})  are smooth since the 
singularities at $A(r)+1=0$  and $A(r)-1=0$  are removable.

It follows from (\ref{eq:4.37}), (\ref{eq:4.38}) that $\{|x|=\alpha_j\}\times\R,j=1,...,m_1,\ 
\{|x|=\beta_k\}\times \R,k=1,...,m_2,$ are boundaries of black or white holes.
In particular if $\alpha_1=\min_{j,k}(\alpha_j,\beta_k)$ and $A(r_1)<-1$
then $r=\alpha_1$ is the boundaries of  a black hole.  
If $\beta_1=\min_{j,k}(\alpha_j,\beta_k)$ and $A(r_1)>1$
then $r=\beta_1$  is the boundary of a white hole.

{\bf Remark 4.4} (Axially symmetric metrics)

Consider an equation (\ref{eq:1.1})  in $\Omega\times\R$  where $\Omega$  is a three-dimensional 
domain.
Let $(r,\theta,\varphi)$  be the spherical coordinates of $x=(x_1,x_2,x_3)$,  i.e.
$x_1=r\sin \theta\cos\varphi,\ x_2=r\sin\theta\sin\varphi,\ 
x_3=r\cos\theta$.

Suppose that $g^{jk}$  are independent of $\varphi:g^{jk}=g^{jk}(r,\theta)$.
Consider characteristic surface $S$ independent of $x_0$  and $\varphi$:
\begin{equation}                                \label{eq:4.39} 
\sum_{j,k=1}^3g^{jk}(r,\theta)\frac{\partial S}{\partial x_j}\frac{\partial S}{\partial x_k}=0,
\end{equation}
where
\begin{equation}                                \label{eq:4.40}
\frac{\partial S}{\partial x_k}=\frac{\partial S}{\partial r}\ \frac{\partial r}{\partial x_k}
+\frac{\partial S}{\partial \theta}\ \frac{\partial\theta}{\partial x_k},\ \ k=1,2,3.
\end{equation}
Substituting (\ref{eq:4.40})  into  (\ref{eq:4.39})  we get an equation:
\begin{equation}                              \label{eq:4.41}
a^{11}(r,\theta)\left(\frac{\partial S}{\partial r}\right)^2+
2a^{12}(r,\theta)\frac{\partial S}{\partial r}\ \frac{\partial S}{\partial\theta}
+a^{22}(r,\theta)\left(\frac{\partial S}{\partial\theta}\right)^2=0.
\end{equation}       
We assume that $a^{jk}(r,\theta)$  are also independent of $\varphi$.
We consider (\ref{eq:4.41})
in two-dimensional domain $\omega$  
such that $\delta_1\leq r\leq \delta_2,\ 0<\delta_3<\theta<\pi-\delta_4$
when $(r,\theta)\in \omega$.
Imposing on $\omega$  and $a^{ij}(r,\theta),\ 1\leq i,j\leq 2,$ 
the conditions of Theorems \ref{theo:3.1} and \ref{theo:4.1}  we shall prove 
the existence of black or white holes in $\Omega\times\R$ with the boundary of the form 
$S_0\times S^1\times \R$  where $\varphi\in S^1,\ x_0\in \R$  and $S_0$  is a closed
Jordan curve in the $(r,\theta)$  plane.

\end{document}